\documentclass[submission,copyright,creativecommons]{eptcs}
\providecommand{\event}{ACL2 2014} % Name of the event you are submitting to
\usepackage{breakurl}             % Not needed if you use pdflatex only.

\usepackage{amssymb} % American mathematical society symbols
\usepackage{graphicx}
\usepackage{fancybox}
\usepackage{url} 
\usepackage{acl2}
\usepackage{float}

\newcommand{\samethanks}[1][\value{footnote}]{\footnotemark[#1]}

\newcommand{\ie}{\mbox{\em i.e.}}
\newcommand{\eg}{\mbox{\em e.g.}}
\newcommand{\viz}{\mbox{\em viz.}}

\title{ Using ACL2 to Verify Loop Pipelining in Behavioral Synthesis}

\author{
Disha Puri\thanks{Dept. of Computer Science, Portland State
  University, Portland OR, 97207. {\tt \{disha,kecheng,xie\}@cs.pdx.edu}}
\and Sandip Ray\thanks{Strategic CAD Labs, Intel
    Corporation, Hillsboro, OR 97124.  {\tt
    sandip.ray@intel.com}}
\and Kecheng Hao\samethanks[1]
\and Fei Xie\samethanks[1]
}

\def\titlerunning{Using ACL2 to Verify Loop Pipelining}
\def\authorrunning{D. Puri, S. Ray, K. Hao \& F. Xie}

\begin{document}

\maketitle

\begin{abstract}
Behavioral synthesis involves compiling an Electronic
System-Level (ESL) design into its Register-Transfer Level
(RTL) implementation.  Loop pipelining is one of the most
critical and complex transformations employed in behavioral
synthesis. Certifying the loop pipelining algorithm is
challenging because there is a huge semantic gap between the
input sequential design and the output pipelined
implementation making it infeasible to verify their
equivalence with automated sequential equivalence checking
techniques. We discuss our ongoing effort using ACL2 to
certify loop pipelining transformation.  The completion of
the proof is work in progress.  However, some of the
insights developed so far may already be of value to the
ACL2 community.  In particular, we discuss the key invariant
we formalized, which is very different from that used in
most pipeline proofs.  We discuss the needs for this
invariant, its formalization in ACL2, and our envisioned
proof using the invariant.  We also discuss some trade-offs,
challenges, and insights developed in course of the project.
\end{abstract}
\section{Introduction}
\label{sec:intro}

Behavioral synthesis is the process of synthesizing an
Electronic System-level (ESL) specification of a hardware
design into an RTL implementation.  The idea of ESL is to
raise the design abstraction by specifying the high-level,
functional behavior of the hardware design.  Designs are
typically specified in a language like C, SystemC, or C++.
The approach is promising since the user is relieved of the
requirement to develop and optimize low-level
implementations.  It has recently received significant
attention, as the steady increase in hardware complexity has
made it increasingly difficult to design high-quality
designs through hand-crafted RTL under aggressive
time-to-market schedules.  Studies have shown that ESL
reduces the design effort by $50\%$ or more while attaining
excellent performance results~\cite{Moussa99}.
Nevertheless, and in spite of availability of several
commercial behavioral synthesis tools, the adoption of the
approach in main-stream hardware development for
microprocessor and SoC design companies has been tentative.
One key reason is the lack of designers' confidence that
the synthesized RTL indeed corresponds to the ESL
specification.  To satisfy the power and performance demands
of modern applications, a behavioral synthesis tool applies
hundreds of transformations, many involving complex and
aggressive optimizations that require subtle and delicate
invariants.  It is unsurprising that the transformations can
contain errors, which can result in bugs in synthesized
hardware.  Thus it is critical to develop mechanized support
to certify the semantic equivalence between ESL and RTL
designs.  On the other hand, the large difference in
abstraction between the two representations makes such
certification non-trivial.

Loop pipelining is a critical transformation in behavioral
synthesis.  The goal of this transformation is to increase
throughout and reduce latency of the synthesized hardware by
allowing temporal overlap of successive loop iterations.  It
is performed by most state-of-the-art commercial synthesis
tools.  Unfortunately, it is also one of the most complex
transformations~\cite{tl:software-popl10}, and creates
challenges in developing viable automated certification
techniques.  In particular, efficient overlapping of loop
iterations creates a design that differs significantly from
its sequential counterpart.  Thus there is little
correspondence in internal signals between the two designs,
making it infeasible to compare them through automated
sequential equivalence checking (SEC).

In this paper, we discuss our ongoing work on using ACL2 for
certifying loop pipelining transformations.  A short
companion paper~\cite{disha-itp14} provides a high-level
summary of our approach suitable for general audience.  In
this paper we provide more ACL2-specific details.
Certification of loop pipelining is a part of our larger
overall effort targeted towards certifying hardware designs
synthesized by behavioral synthesis.  An explicit goal of
this overall framework is to make use of fully automatic
decision procedures to the extent possible.  Many components
of the project are quite mature, but they are based on
automatic sequential equivalence checking (SEC).  The work
discussed here is the key theorem proving component.  We
discuss the overall framework and how our work fits with the
rest of the components in Section~\ref{sec:background}.

Our ACL2 proof is ongoing, and its completion will still
require significant work.  Nevertheless, our work so far has
already produced insights that we believe are worth
disseminating to the community.  The paper touches upon the
following three points.

First, a key technical contribution is the invariant we defined.  It
differs from a typical invariant used for correctness of pipelined
systems in that it explicitly specifies the correspondence between the
sequential and pipelined programs at each transition.  We elaborate on
our definition in Section~\ref{sec:proof}.

Second, our project is somewhat different from the
traditional applications of ACL2 in hardware verification.
First, since an over-arching goal is to exploit automatic
decision procedures, we use theorem proving primarily to
complement automated tools.  Second, we eschew theorem
proving on inherently complex or low-level implementations.
Third, interactive theorem proving is acceptable for
one-time use, \eg\  in certification of a single
transformation, but not as part of a methodology that
requires ongoing use in certification of each design.  The
constraints are imposed by the the environment in which we
envision our framework being deployed: it may not be
possible to have a dedicated team of experts doing theorem
proving as full-time job.  Finally, as discussed above, our
approach is targeted towards verification of procedures
which are closed-source (and exceedingly complex), thus
making traditional program verification techniques unusable.
We believe our approach shows a novel way in which theorem
proving can be applied even under those constraints, in
concert with automatic SEC.

Finally, we comment upon our experience in a new user's
experience with ACL2.  ACL2, like any other interactive
reasoning tool, has a significant learning curve.
Unfortunately, most of the participants of the team in this
task had no previous exposure to automated theorem proving.
In particular, the majority of the work in this effort was
done by the first author and it represents her first project
in formal methods beyond exercises in the ACL2 textbook.
Our experience provides some insight in the complexity of
developing industrial-scale solutions with ACL2 from a new
user's perspective.

The remainder of the paper is organized as
follows. Section~\ref{sec:background} provides some
background on the overall project and explains the context
of our theorem proving work.  In Section~\ref{sec:ccdfg} we
discuss our formalization of {\em CCDFG}, a structure which
formalizes intermediate representations used in behavioral
synthesis.  Section~\ref{sec:formalization} discusses our
correctness statement, while Section~\ref{sec:proof}
presents the key proof steps, and the definition of our
invariant.  In Section~\ref{sec:user-experience} we share
some insights we gleaned from this experience on difficulty
of using ACL2 in a proof project of this scale for a new
user.

 \section{Background and Context}
\label{sec:background}

In this section, we discuss the overall framework of the project we
are working on, and how the certification of loop pipelining fits into
this project.  The description here is, of course, necessarily
minimal.  The reader interested in a thorough understanding of other
components of the project is welcome to review the prior
publications~\cite{rhcxy:atva-09,hxry:date-10}.

\subsection{Behavioral Synthesis}

We start with a brief summary of behavioral synthesis, highlighting
the aspects necessary to understand the rest of our presentation.
Note that behavioral synthesis is a complex technology whose
description is far beyond the scope of this paper.  However, there are
several behavioral synthesis tools available, including both
commercial and academic ones, with extensive manuals, documentations
and tutorials providing user-level description of their
workings~\cite{spark,xpilot,legup}, and we encourage the
interested reader to these materials for further information.  

At a high level, a behavioral synthesis tool can be best viewed as a
``compiler'' that takes an ESL description and generates RTL.
Analogous to a regular compiler, a behavioral synthesis tool performs
the standard lexical and syntax analysis to generate in intermediate
representation (IR).  The IR is then subjected to a number of
transformation, which can be categorized into three phases.

 \begin{itemize}
\item {\bf Compiler Transformations:} These include typical
  compiler operations, {\em e.g.,} dead-code elimination,
  constant propagation, loop unrolling, etc.  A design may
  undergo hundreds of compiler transformations.
\item {\bf Scheduling Transformations:} Scheduling entails
  computing for each operation the clock cycle of its
  execution, accounting for hardware resource constraints
  and control/data flow.  Loop pipelining, the subject of
  this paper, is a component of this phase. 
\item {\bf Resource Allocation and Control Synthesis:} This phase
  involves mapping a hardware resource to each operation (\eg, the
  ``{\tt +}'' operation may be mapped to a hardware adder), allocating
  registers to variables, and generating a controlling finite-state
  machine to implement the schedule.
\end{itemize}
After the three phases above, the design can be expressed in
RTL.  The synthesized RTL may be subjected to further manual
tweaks to optimize for area, power, etc.

\subsection{Certification Framework}

The overall goal of our project is to provide a mechanized framework
for certifying hardware designs synthesized from ESL specifications by
commercial behavioral synthesis tools.  One obvious approach, of
course, is to apply standard verification techniques (\eg, SEC or
theorem proving) on the {\em synthesized RTL} itself.  Unfortunately,
such a methodology is not practical.  The large gap in abstraction
between the ESL and RTL descriptions means that there is little
correspondence in internal variables between the two.  Consequently,
direct SEC between the two reduces to cost-prohibitive computation of
input-output equivalence.  On the other side, applying theorem proving
is also troublesome since extensive manual effort is necessary and
this effort needs to be replicated for each different synthesized
design.  On the opposite end of this, it is also infeasible to
directly certify the implementation of the {\em synthesis tool} via
theorem proving.  In addition to being highly complex and thus
potentially requiring prohibitive effort to formally verify with any
theorem prover, the implementations are typically closed-source and
closely guarded by EDA vendors and thus out of reach of external
automated reasoning communities.  

To address this problem, previous work developed two key SEC
solutions, which we will refer to below as {\em Back-end} and {\em
  Front-end}.  We then discuss the gap between them, which is being
filled by theorem proving efforts.

\medskip

\noindent {\bf Back-end SEC:} The key insight behind
back-end SEC is that automated SEC techniques, while
ineffective for directly comparing synthesized RTL with the
top-level ESL description, is actually suitable to compare
the RTL with the intermediate representation (IR) generated
by the tools after the high-level (compiler and scheduling)
transformations have been applied.  In particular,
operation-to-resource mappings generated by the synthesis
tool provide the requisite correspondence between internal
variables of the IR and RTL.  Furthermore, a key insight is
that while the implementations of transformations are
unavailable for commercial EDA tools, most tools provide
these IRs after each transformation application together
with some other auxiliary information.  To exploit these, an
SEC algorithm was developed between the IR (extracted from
synthesis tool flow after these transformations) and
RTL~\cite{rhcxy:atva-09,hxry:date-10}.  The approach scales
to tens of thousands of lines of synthesized RTL.

\medskip

\noindent {\bf Front-end SEC:} Of course the back-end SEC
above is only meaningful if we can certify that the input
ESL indeed corresponds to the extracted IR produced after
the compiler and scheduling transformations applied in the
first two phases of synthesis. To address this, another SEC
technique was developed to compare two IRs.  The idea then
is to obtain the sequence of intermediate representations
$\mbox{IR}_0,\ldots,\mbox{IR}_n$ generated by the compiler
and scheduling transformations, and compare each pair of
consecutive IRs with this new algorithm.  Then back-end SEC
can be used to compare $\mbox{IR}_n$ with the synthesized
RTL, completing the flow.

\bigskip

\noindent {\bf A Methodology Gap:} Unfortunately, the front-end SEC
algorithm can only compare two IRs that are structurally close.  If a
transformation significantly transforms the structure of an IR then
the heuristics for detecting corresponding variables between the two
IRs will not succeed, causing equivalence checking to fail.
Unfortunately, loop pipelining falls in the category of
transformations that significantly changes the structure of the IR.
It is a quintessential transformation that changes the control/data
flow and introduces additional control structures (\eg, to eliminate
hazards).  This makes front-end SEC infeasible for its certification.
On the other hand, it is also a complex and error-prone
transformation, and ubiquitous across different behavioral synthesis
tools. Furthermore, most commercial implementations are of course
proprietary and consequently not available to us for review; applying
theorem proving on those implementations is not viable from a
methodology perspective.  Thus a specialized approach is warranted for
handling its certification.

\subsection{A Reference Pipeline Approach}
\label{sec:pipelining-challenges}

To develop a specialized approach for pipelines, our key observation
is that while the transformation {\em implementation} is inaccessible
to us, commercial synthesis tools typically generates a report
specifying pipeline parameters (\eg, pipeline interval, number of loop
iterations pipelined, etc.).  Our approach then is to develop an
algorithm that takes as inputs these parameters and an IR ${\cal{C}}$
for the design before pipelining, and generates a {\em reference
  pipelined IR} ${\cal{P}}$.  Note that our algorithm is much simpler
than that employed during synthesis; while the former includes
advanced heuristics to {\em compute} pipeline parameters (like
pipeline interval, number of iterations pipelined etc.), we merely use
the values provided by its report.  To certify a synthesized RTL with
pipelines, it is sufficient to (1)~check that {\em our algorithm} can
generate a pipeline ${\cal{P}}$ for the parameters reported by
synthesis, (2)~use SEC to compare ${\cal{P}}$ with the synthesized
RTL, and (3)~prove (using theorem proving) the correctness of our
algorithm.

Previous work~\cite{hrx:dac-12} in fact justified the viability of
steps 1 and 2 above; such a reference pipeline generation algorithm
was developed and used to successfully compare a variety of pipelined
designs across various application domains.  
%% Comment: how difficult are the SEC problems it generates?
%%
This suggested that the
approach of using a reference implementation is viable for certifying
industrial synthesized pipelines.  However, a key (and perhaps the
most complicated) component of the approach, step 3, was missing.  The
algorithm was not verified (indeed, not implemented in a formal
language), rendering the ``certification'' flow unsound.

The unsoundness mentioned above is not just an academic notion.  In
fact, merely by going through the formalization process and thinking
about necessary invariants, we have already found multiple bugs in
this algorithm.  Thus it is critical to develop a mechanized proof of
correctness of this implementation.  Unfortunately, it is not easy to
verify the original pipeline generation algorithm as written.  Its
author was an expert in behavioral synthesis but not in program
verification or theorem proving; consequently, the algorithm, while
simpler than that implemented in a synthesis tool, was still a highly
complex piece of code.  In particular, since it was not written with
correctness in mind, it is difficult to decompose the algorithm into
manageable pieces with nice invariants.

One way to address this problem is to ``buckle down'' and
verify the pipeline generation algorithm (and fixing the
bugs found in the process).  However, a key insight in our
case is that we can get away without verifying such a
complex implementation.  After all, there is nothing
``sacred'' about this specific algorithm for pipeline
generation: given the steps described above, {\em any}
verifiable pipeline generation algorithm would
suffice.\footnote{Note that our algorithm {\em must} create
  a pipeline in accordance with the pipeline parameters
  obtained from the behavioral synthesis tools else we may
  fail to certify correct designs. However, in practice, we
  have not found this to be a problem.}  Thus the approach
of this paper can be viewed as a rational deconstruction of
the pipeline synthesis algorithm of the previous work.  We
identify the key invariant that we need to maintain for
proving computational equivalence between the pipelined and
un-pipelined loops and design the algorithm to explicitly
maintain that invariant.

\section{CCDFG}
\label{sec:ccdfg}

In order to formalize and prove the correspondence between pipelined
and unpipelined IRs, a first step is to define a formalization of the
IRs themselves.  We call our formalization of IRs {\em Clocked Control
  Data Flow Graph} (CCDFG).   An informal description of CCDFG has been
provided before~\cite{rhcxy:atva-09}. It can be best viewed as a
traditional control/data flow graph used by most compilers, augmented
with a schedule. Control flow is broken into basic blocks.
Instructions are grouped into microsteps which can be executed
concurrently.  A scheduling step is a group of microsteps which can be
executed in a single clock cycle. The state of a CCDFG at a particular
microstep is a list of all the variables of a CCDFG with their
corresponding values.

The semantics of CCDFG require a formalization of the
underlying language used to represent the individual
instructions in each scheduling step.  The underlying
language we use is the LLVM~\cite{llvm}.  It is a popular
compiler infrastructure for many behavioral synthesis tools
and includes an assembly language front-end.  At the point
of this writing we support a limited subset of LLVM, which
however is sufficient to handle all the designs we have
seen.  Instructions supported include assignment, load,
store, bounded arithmetic, bit vectors, arrays, and pointer
manipulation instructions.  We define the syntax of each
type of statement by defining an ACL2 predicate.  For
example, in our syntax, an assignment statement can be
expressed as a list of a list of a variable and an
expression.

\small
\begin{verbatim}
(defun assignment-statement-p (x)
  (and (equal (len x) 1)
       (and (equal (len (car x)) 2)
            (first (car x)) (symbolp (first (car x)))
            (expression-p (second (car x))))))
\end{verbatim}
\normalsize

Here, an expression can further be of multiple types, \eg, load
expression (loading the value of a variable from memory), add
expression (add two variables), xor expression (xor of two variables)
etc., where each expression includes the operation applied to the
appropriate number of arguments.

\small
\begin{verbatim}
(defun add-expression-p (x)
  (and (equal (len x) 3)
       (equal (first x)  'add)
       (variable-or-numberp (second x))
       (variable-or-numberp (third x))))

(defun expression-p (x)
  (and (consp x)
       (or (load-expression-p x) (add-expression-p x) (xor-expression-p x)...)))
\end{verbatim}
\normalsize

We provide semantics to these instructions through a
state-based operational formalization as is common with
ACL2.  We define the notion of a CCDFG state, which includes
the states of the variables, memory, pointers, etc.  Then we
define the semantics of each instruction by specifying how
it changes the state.  Thus, for assignment statement we
will have a function {\tt execute-assignment} that specifies
the effect of executing the assignment statement on a CCDFG
state.

Defining the semantics of most supported statements is
straightforward, with one exception.  The exception is the
so-called ``$\phi$-construct'' available in LLVM.  A
$\phi$-construct is a list of $\phi$-statements.  A
$\phi$-statement is $v := \phi [\sigma, bb1] [\tau, bb2]$,
where $v$ is a variable, $\sigma$ and $\tau$ are
expressions, and $bb1$ and $bb2$ are basic blocks: if while execution, 
the previous scheduling step of the $\phi$-statement is
$bb1$ then it is the same as the assignment
statement $v := \sigma$; if reached from $bb2$ {\em i.e.}, the previous scheduling step is $bb2$, 
it is the
same as $v := \tau$; the meaning is undefined otherwise. The
construct is complex since the effect of executing this
statement on a CCDFG state $s$ depends not only on the state
$s$ but also on how $s$ is reached by the control flow.
Unfortunately, $\phi$-statements abound in pipelined (and
sequential) designs we have worked with -- they are used to
find the value of loop dependencies.  Consequently, the complexity induced
by this instruction cannot be avoided.

\small
\begin{verbatim}
(defun phi-expression-p (x)
  (and (consp x) (equal (len x) 1)
       (consp (car x)) (> (len (car x)) 2)
       (equal (caar x) 'phi) (phi-l (cdr (car x)))))

(defun phi-statement-p (x)
  (and (consp x) (equal (len x) 2)
       (symbolp (first x)) (first x)
       (phi-expression-p (cdr x))))
\end{verbatim}
\normalsize

Here {\tt phi-1} recognizes an expression of the form {\tt
  ((E0 b) (E1 b-prime))} where {\tt E0} and {\tt E1} are
expressions and {\tt b} and {\tt b-prime} are symbols
representing basic blocks.  Thus in ACL2, the
$\phi$-statement looks like {\tt (v (phi ((E0 b) (E1
  b-prime))))}.  Finally, the execution semantics requires
the additional parameter {\tt prev-bb} to track the previous
basic block.

%% Comment: Function definition for execute-phi uses the functions
%%evaluate-val, replace-val and variables-of which are not
%%explained.

\small
\begin{verbatim}
(defun choose (choices prev-bb)
  (if (or (equal (nth 1 (first choices)) prev-bb)
          (equal (symbol-name (nth 1 (first choices))) prev-bb))
      (nth 0 (first choices))
    (nth 0 (second choices))))
    
(defun evaluate-val (val bindings)
  (if (symbolp val) 
      (cdr (assoc-equal val bindings)) 
    val))

(defun execute-phi (stmt init-state prev-bb)
  (let* ((expr (cdr stmt))
         (var (first stmt))
         (val (evaluate-val (choose (cdr (car expr)) prev-bb) (car init-state))))
    (list (replace-var var val (variables-of init-state)) 
          (memory-of init-state) 
          (pointers-of init-state))))         
\end{verbatim}
\normalsize

%% Added
The {\em init-state} represents the state of a CCDFG before executing $\phi$-statement. The function 
{\em variables-of} is used to get a list of all the variables of {\em init-state} with their corresponding values. 
{\em replace-var} replaces the values of the variable {var} to {val} in the list of those variables.

\section{Correctness of Loop Pipelining}
\label{sec:formalization}

For the purposes of this paper, a {\em pipelinable loop} is
a loop with the following restrictions~\cite{hrx:dac-12}:
\begin{enumerate}
\item no nested loop;
\item only one $Entry$ and one $Exit$ block; and
\item no branching between the scheduling steps.
\end{enumerate}
These restrictions are not meant to simplify the problem, but reflect
the kind of loops that can be actually pipelined during behavioral
synthesis.  For instance, synthesis tools typically require inner
loops to have been fully unrolled (perhaps by a previous compiler
transformation) in order to pipeline the outer loop.

\begin{figure}
\begin{center}
\begin{tabular}{ccc}
\includegraphics[height=1.8in]{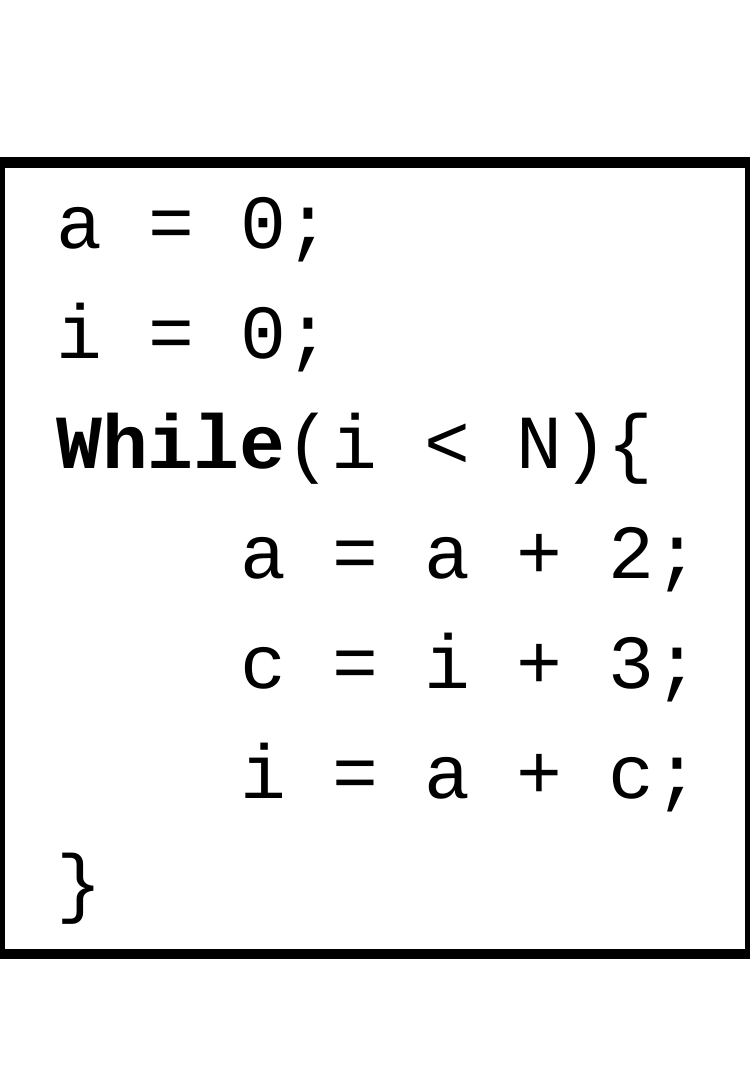}
& \hspace{1cm}
\includegraphics[height=1.8in]{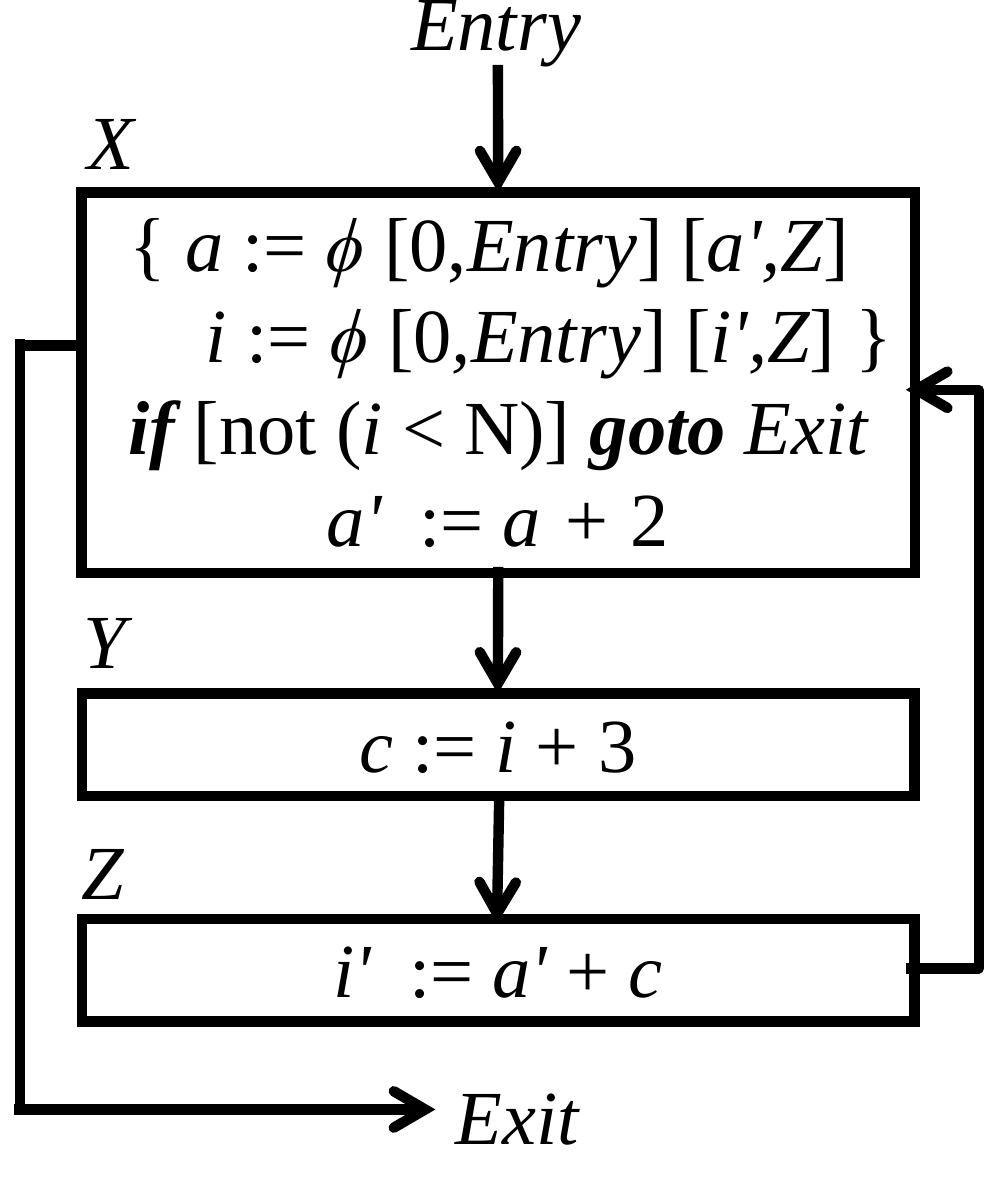}
& \hspace{1cm}
\includegraphics[height=1.8in]{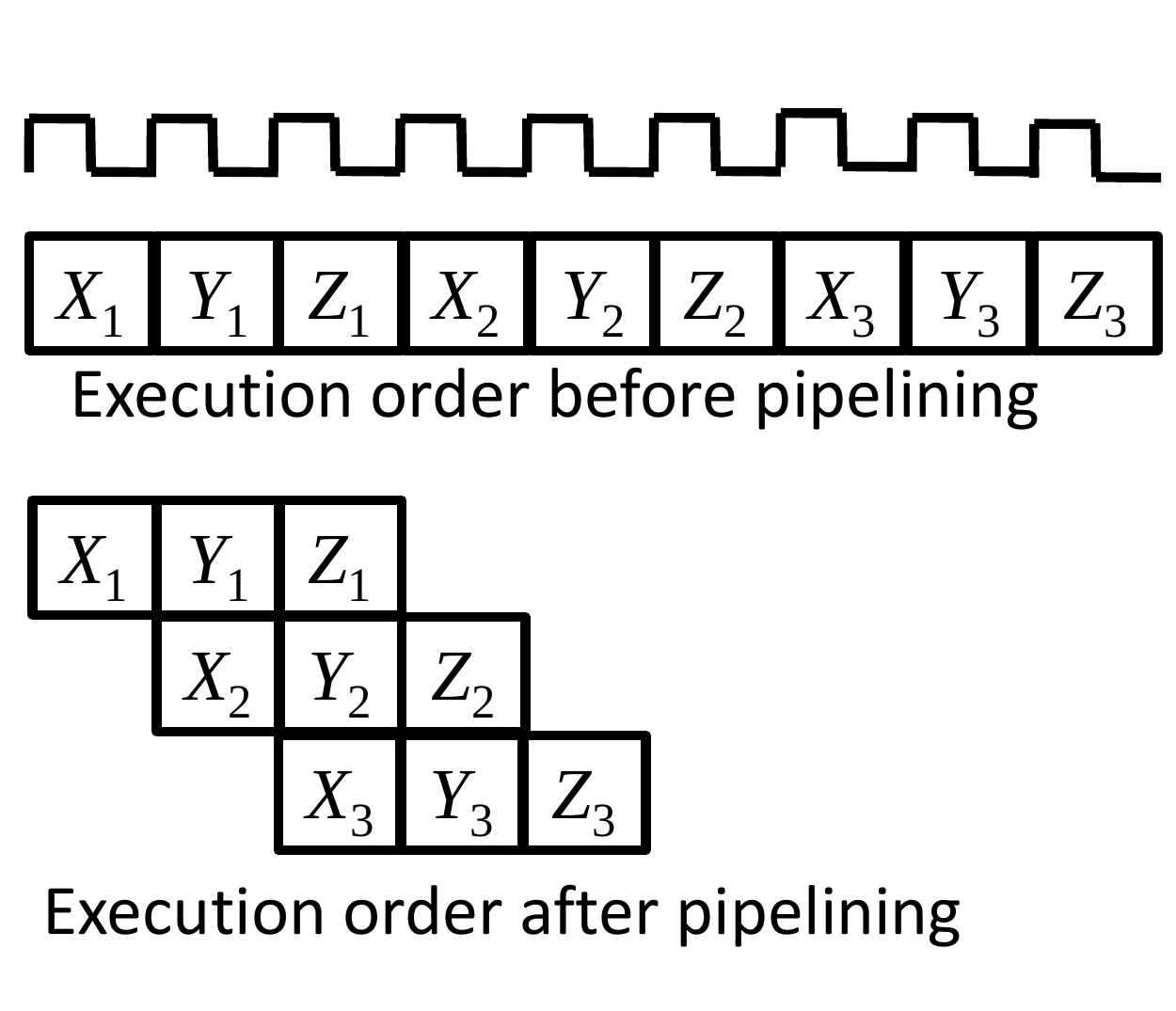}
\\
(a) & \hspace{1cm} (b) & \hspace{1cm} (c)
\end{tabular}
\end{center}
\caption{(a) Loop in C (b) Loop CCDFG before pipelining (c) Pipelining increases throughput}
\label{fig:high-level-synthesis}
\end{figure}

%%Comment:  it would be nice to have
%%a more complete description of what the Clocked Control/Data Flow
%%Graph (CCDFG) illustrations mean, e.g., what does each edge between
%%boxes signify.
%% Comment: quoted variables
%% Comment: Figure 1(c) is not referenced in the
%%paper, leaving this reviewer with a question because the point seems
%%to (partially) be that it's not necessarily simple to just "unroll"
%%the initial specification (program) into some kind of parallel
%%(hardware) design. 

Figure~\ref{fig:high-level-synthesis}(a) illustrates the C
code (ESL description) for a loop.  The C code does not have
a schedule or the concept of a clock cycle.
Figure~\ref{fig:high-level-synthesis}(b) shows CCDFG of the
sequential loop just before loop pipelining. The loop has
three scheduling steps: $X$, $Y$ and $Z$.  The scheduling
step before the loop is $Entry$ and after the loop is
$Exit$.
%% Added 
The edges  in the CCDFG indicate the control flow.
Note that the sequential CCDFG has SSA structure, as a result variable $a$ and $i$ are not assigned 
more than once and we require the quoted variables $a'$ and $i'$.
Note that there is a $\phi$-statement in the
first scheduling step of the loop.  This $\phi$-statement
accounts for variables whose values are different based on
the first time we enter the loop. 

%% Added
Assuming that each scheduling step takes one clock cycle to complete,
one iteration of the sequential design takes three
clock cycles and three iterations would take nine clock cycles.
Given a pipeline interval of one, three iterations of the pipelined loop takes only five clock cycles as shown in
Figure~\ref{fig:high-level-synthesis}(c).
So, loop pipelining is important to reduce latency
and improve throughput of the synthesized pipeline circuit.  
The main lemma involved in the correspondence proof between
the sequential and pipelined CCDFG can be stated in ACL2 as
follows.\footnote{Although we state this lemma as a {\tt
    defthm} in the paper, as we mentioned before we have not
  completed its proof in ACL2.  We do have a sufficient
  proof sketch.  However, the final form of this theorem,
  when proven, can have some differences, \eg, additional
  hypotheses that we have overlooked so far.}

\begin{verbatim}
(defthm correctness-statement-key-lemma
  (implies (and (posp k) (posp pp-interval) (posp m)
                (equal pp-ccdfg (superstep-construction pre loop pp-interval m))
                (not (equal pp-ccdfg "error")))
           (equal 
            (in-order (get-real (run-ccdfg (first pp-ccdfg) (second pp-ccdfg) 
                                           (third pp-ccdfg) k init-state prev)))
            (in-order (run-ccdfg pre loop nil (+ (- k 1) (ceil m pp-interval)) 
                                 init-state prev)))))
\end{verbatim}

The theorem involves several ACL2 functions, \eg, {\tt
  get-real}, {\tt superstep-construction}, etc.  We do not
discuss the detailed definitions of these functions in the
paper, but they are available with the supporting ACL2
script for this workshop.  We provide a brief, informal
description of some of the critical functions in the theorem
below.  Following is an English paraphrase of the theorem.

\begin{quote}
 If the pipeline generation succeeds without error,
 executing the pipelined CCDFG loop for $k$ iterations
 generates the same state of the relevant variables as
 executing the sequential CCDFG for some $k'$ iterations.
 The explicit value of $k'$ is given by the term {\tt (+ (-
   k 1) (ceil m pp-interval))}.
 \end{quote}

\noindent
Two key functions that appear in the theorem above are {\tt
  superstep-construction} and {\tt run-ccdfg}.  The
functrion {\tt superstep-construction} combines the
scheduling steps of successive iterations to create the
``scheduling supersteps'' of pipelined CCDFG.  If there are
data-hazards and pipelined CCDFG cannot be generated as per
the pp-interval given, the function generates an ``error''.
The function {\tt run-ccdfg} runs a CCDFG including a
pipelinable loop in three parts, first the prologue before the
loop, next the loop itself, and finally the epilogue past the
loop.\footnote{Of course one can have the standard function
  {\tt run} that executes the entire CCDFG rather than in
  parts.  However, for reasons that will be clear when we
  define the invariant, in our case it is easier to do most
  of the work with the execution in three parts and then
  assemble them into a final theorem about the CCDFG run in
  the end.}  This function is defined as follows, where {\tt
  prefix} determines the previous scheduling step of the
iteration (required to resolve $\phi$-statements).

\begin{verbatim}
(defun run-ccdfg (pre loop post iterations init-state prev)
  (let* ((state1 (run-block-set pre init-state nil prev))
         (state2 (run-blocks-iters loop state1 iterations (prefix loop)))
         (state3 (run-block-set post state2 nil (prefix post)))
    state3))
\end{verbatim}
Finally, the function {\tt get-real} removes from the
pipelined CCDFG state, all auxiliary variables introduced by
the pipeline generation algorithm itself, leaving only the
variables that correspond to the sequential
CCDFG,\footnote{The algorithm has to introduce new variables
  in order to eliminate hazards.  One consequence of this is
  that the new variables so introduced must not conflict
  with any variable subsequently used in the CCDFG.  Since
  we do not have a way to ensure generation of fresh
  variables, this constraint has to be imposed in the
  hypothesis.}  and {\tt in-order} normalizes ``sorts'' the
components in a CCDFG state in a normal form so that the
sequential and pipelined CCDFG states can be compared with
{\tt equal}.

\section{Algorithm, Invariant, and Proof}
\label{sec:proof}

\begin{figure}
\begin{center}
\includegraphics[height=2.2in] {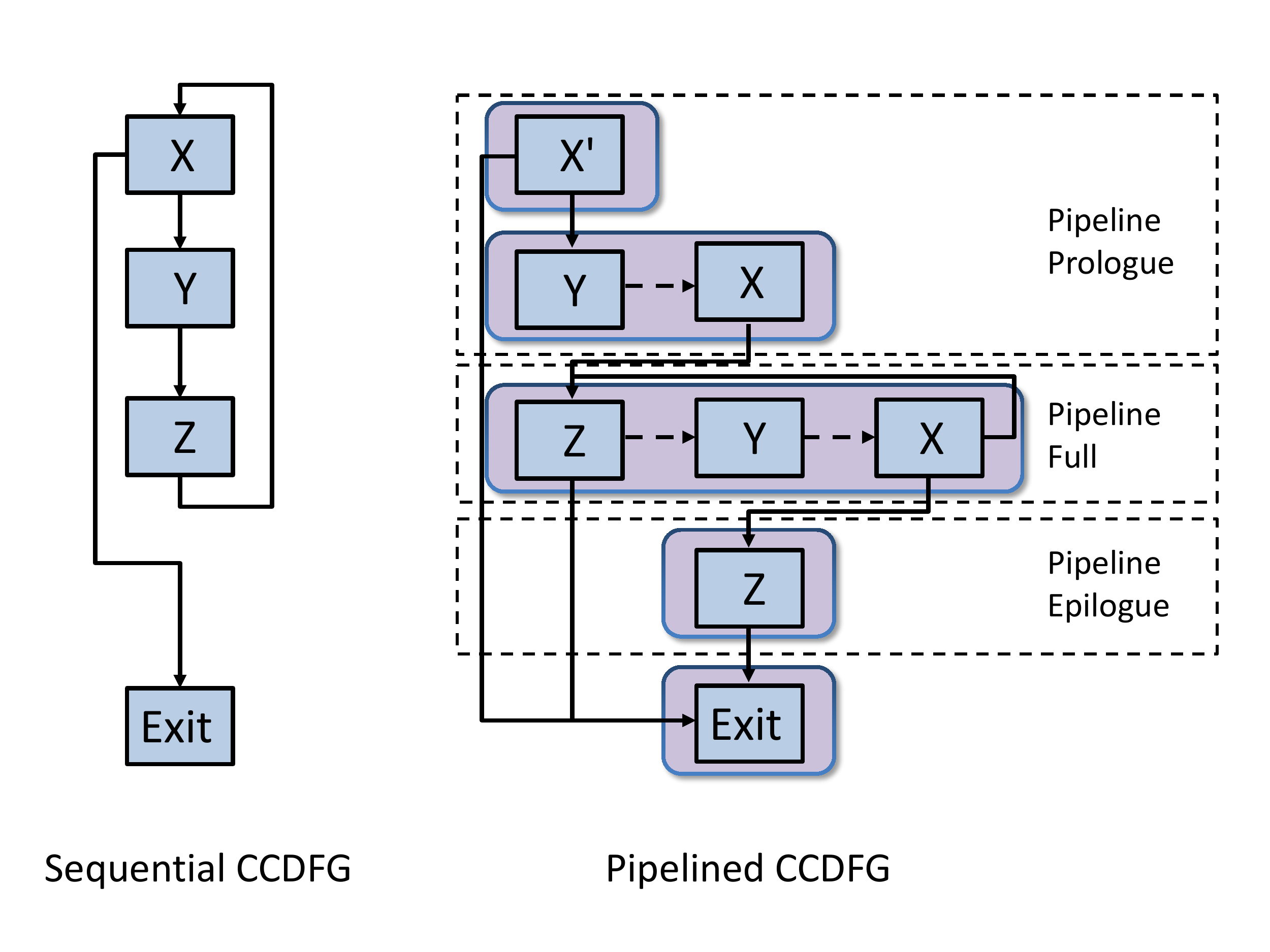}
\end{center}
\caption{Conversion of sequential CCDFG to pipelined CCDFG
  using superstep construction. Pipeline full stage depicts
  the loop in pipelined CCDFG. Horizontal dashed arrows indicate
  data forwarding.}
\label{fig:invariant-intro}
\end{figure}

Pipeline synthesis is based on the key observation that
execution of successive iterations can be overlapped as long
as no data hazard is introduced.  Thus, the two main
activities of a pipeline synthesis algorithm are to
(1)~identify possible hazards, and (2)~eliminate them,
typically by introduction of new variables.  In our case,
the identification is simplified since the synthesis tool
provides a pipeline interval; thus, instead of {\em
  discovering} a pipeline interval ourselves so that no
hazard is introduced, we merely need to work with the
provided interval.

\bigskip

\noindent {\bf Remark:} In addition to (1) and (2) above, we
need another operation to eliminate the thorny
$\phi$-statement.  The insight behind this procedure is that
the $\phi$-statement is used to evaluate values of variables based on 
the previous basic block; so if
the loop is unwound by one iteration the statement can be
replaced by corresponding assignment statements (\eg, for
the first (unwound) iteration, the assignment statement
would correspond to the component of $\phi$ in the {\tt
  Entry} block, and in the rest it would correspond to the
component of $\phi$ for the last basic block in the body of
the loop).  The correctness of $\phi$-elimination is
independent of the remainder of the pipeline operations.
Once this is performed on CCDFG $C$, we have a sequential
CCDFG $C'$ with no $\phi$-statement but whose loop includes
a preamble that corresponds to the first iteration of the
loop in $C$ and whose ``loop body'' corresponds to the
remaining loop iterations in $C$.

\bigskip

There are two key operations involved in pipeline synthesis
: (1)~insertion of shadow variables, and (2)~combining
microsteps from different iterations of the sequential loop
into one ``superstep'' of the pipelined CCDFG.  Inserting
shadow variables protects variables that are written and
then read within the same iteration from introducing hazards
when iterations are overlapped.  To understand its role,
suppose a variable {\tt x} is first written and then read in
the same iteration and consider overlapping two iterations.
Then the write from the second iteration may overwrite the
value of {\tt x} before the read from the first iteration
had a ``chance'' to read the previous value.  To avoid this,
we introduce a fresh variable {\tt x\_reg}, which preserves
the old value of {\tt x}, and reads to {\tt x} are replaced
by reads to {\tt x\_reg}.  Finally, the superstep
construction entails combining scheduling steps of
successive iterations (the number of overlapping iterations
being defined by the pipeline interval provided).  We
combine two corresponding scheduling steps from successive
iterations if there is no hazard introduced, \viz, if a
variable is written in scheduling step $S$ and read
subsequently in $S'$ then $S'$ cannot be in a superstep that
precedes $S$.  $S$ and $S'$ can be in a single superstep
since we implement data forwarding.  Note that the superstep
consruction simply fails if we cannot combine the requisite
scheduling steps.  So it is possible that although a
pipelined CCDFG exists with the provided pipeline interval
but we fail to construct it by being conservative in our
analysis of potential hazards.  However, all our soundness
claims are only for pipelines that we can synthesize.

Our key invariant defines a ``correspondence relation''
between the backedges of the sequential and pipelined CCDFG.
The relation can be informally paraphrased as
follows~\cite{disha-itp14}.

\begin{quote}
Let $S$ be a sequential loop and $G$ be the pipelined loop
generated from our algorithm. The pipelined loop consists of
three stages as depicted in
Figure~\ref{fig:invariant-intro}: prologue $G_p$, full stage
$G_l$, and epilogue $G_e$.  Let $s_l$ be any state of $G$
poised to execute $G_l$, and let $k$ be any number such that
the loop of $G$ is not exited in $k$ iterations from $s_l$.
Then executing $G_p$ followed by $k$ iterations of $G_l$ is
equivalent to executing first iteration of $S$, say $S_1$
followed by $(k - 1)$ iterations of $S$ together with a
collection of ``partially completed'' iterations of
$S$.\footnote{The formalization actually characterizes each
  incomplete iteration, \eg, if the pipeline includes $d$
  iterations and successive iterations are introduced in
  consecutive clock cycles, then the $i$-th iteration has $i
  - 1$ incomplete scheduling steps.}
\end{quote}

Suppose the number of scheduling steps of the first iteration in pipeline
prologue is {\tt m} and the pipeline interval is {\tt i}. We
calculate the value of {\tt m} based on the number of scheduling steps in a CCDFG
and the pipeline interval. In our example, {\tt m} is $2$
and {\tt i} is $1$.  The invariant implies that starting
from the same initial state, executing $G_p$ and {\tt k}
iterations of $G_l$ is the same as executing {\em k}
iterations of $S$, followed by {\tt m} scheduling steps of
$S$, followed by {\tt (m - i)} scheduling steps of $S$, by
{\tt (m - 2i)} scheduling steps of $S$, etc. till the number
remains positive.  The invariant can be defined in ACL2 as
follows.

\small
\begin{verbatim}
(defun get-m-blocks-seq (m seq-loop pp-interval)
  (if (or (not (posp m))
          (not (posp pp-interval))) nil
    (if (<= m pp-interval) 
        (take-n m seq-loop) ; get first m blocks of loop
        (append (take-n m loop)
              (get-m-blocks-seq (- m pp-interval) seq-loop pp-interval)))))


(defun pipeline-loop-invariant 
       (pp-state k seq-pre seq-loop pp-interval init-state prev m)
   (let* ((seq-loop-top (run-block-set seq-pre init-state nil prev))
          (seq-loop-k (run-blocks-iters seq-loop seq-loop-top (- k 1) prev))
          (seq-loop-x (run-block-set (get-m-blocks-seq m seq-loop pp-interval) 
                                     seq-loop-k nil prev)))
     (equal pp-state seq-loop-x)))
            
(defthm invariant-holds
  (implies (and (posp k) (posp m) (posp pp-interval)
                (equal pp-ccdfg 
                       (superstep-construction seq-pre seq-loop pp-interval m))
                (not (equal pp-ccdfg ``error'')))
           (pipeline-loop-invariant 
             (run-ccdfg-k (car pp-ccdfg) (second pp-ccdfg) 
                           k init-state prev)
             k seq-pre seq-loop pp-interval init-state prev m))
\end{verbatim}
\normalsize

Here, {\tt seq-loop-k} represents the state after executing
{\tt seq-loop} for {\tt k} iterations starting from state
{\tt init-state}, {\tt seq-loop-x} represents the state
after executing the blocks defined by function {\tt
  get-m-blocks-seq} on the on the sequential loop {\tt
  seq-loop} starting from state {\tt seq-loop-k}.  The
invariant then says that the state thus obtained is equal to
the input {\em pp-state}.  Note that {\tt pp-state} is the
state obtained by function {\tt run-ccdfg-k} which executes
pipeline prologue and {\tt k} iterations of pipeline full
stage starting from state {\tt init-state}.
 
\medskip 

As is standard with proofs involving invariants, there are
two obligations to prove the correctness (in this case the
main lemma of the previous section - 
% Added
{\em correctness-statement-key-lemma}), 
\viz, that it is indeed
an invariant, and that its invariance is sufficient to imply
the desired correctness theorem.  Here we give a sense of
our envisioned proof.

\begin{figure}
\begin{center}
\includegraphics[height=2.2in]{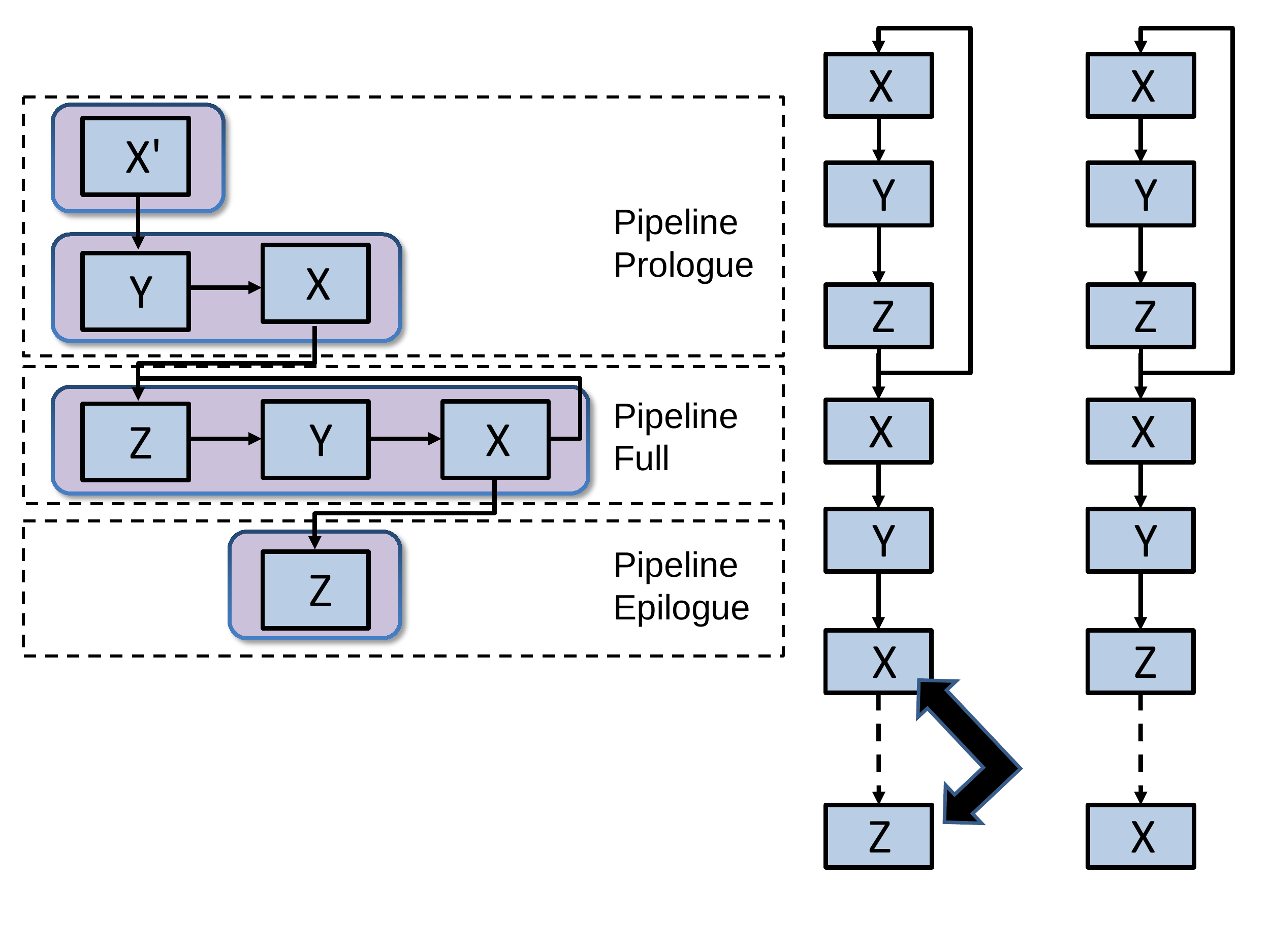}
\end{center}
\caption{Correctness of invariant implies the correctness statement} 
\label{fig:invariant-implies-correctness}
\end{figure}

Our invariant is defined specifically to make the proof of
sufficiency straightforward.  Suppose the $P$ is the
pipelined CCDFG and $S$ is the sequential CCDFG.
Equivalence of CCDFG states of $P$ and $S$ follows from the
invariant by noting that the epilogue $P_e$ exactly
constitutes the incomplete scheduling steps of $S$ specified
by the invariant
(cf. Figure~\ref{fig:invariant-implies-correctness}).

The proof of invariance of this predicate is, of course, the
main ``work horse'' in this exercise.  The proof depends on
a fundamental idea for pipelining, \viz, commutability of
independent instructions.

\begin{quote}
Suppose that the
set of variables written and read by two consecutive
operations $a$ and $b$ is disjoint.  Then executing $a$
followed by $b$ generates the same result as executing $b$
followed by $a$.
\end{quote}

If we view the scheduling steps in
Figure~\ref{fig:high-level-synthesis} as arranged in a
matrix, then the sequential execution proceeds column-wise
along the matrix while the pipelined execution proceeds
row-wise.  Thus the core proof obligation involves the
following two proof requirements. 

\begin{itemize}
\item Our pipelining algorithm correctly combines the
  ``appropriate'' scheduling supersteps which do not have
  read-write hazards.
\item Given that there are no read-write hazards at
  appropriate places, executing scheduling steps row-wise is
  same as executing those scheduling steps column-wise in
  the pipelined CCDFG.  This makes use of the fundamental
  observation above.
\end{itemize}

\begin{figure}
\begin{center}
\begin{tabular}{c@{\hspace*{.3mm}}c}
\includegraphics[width=3.5in]{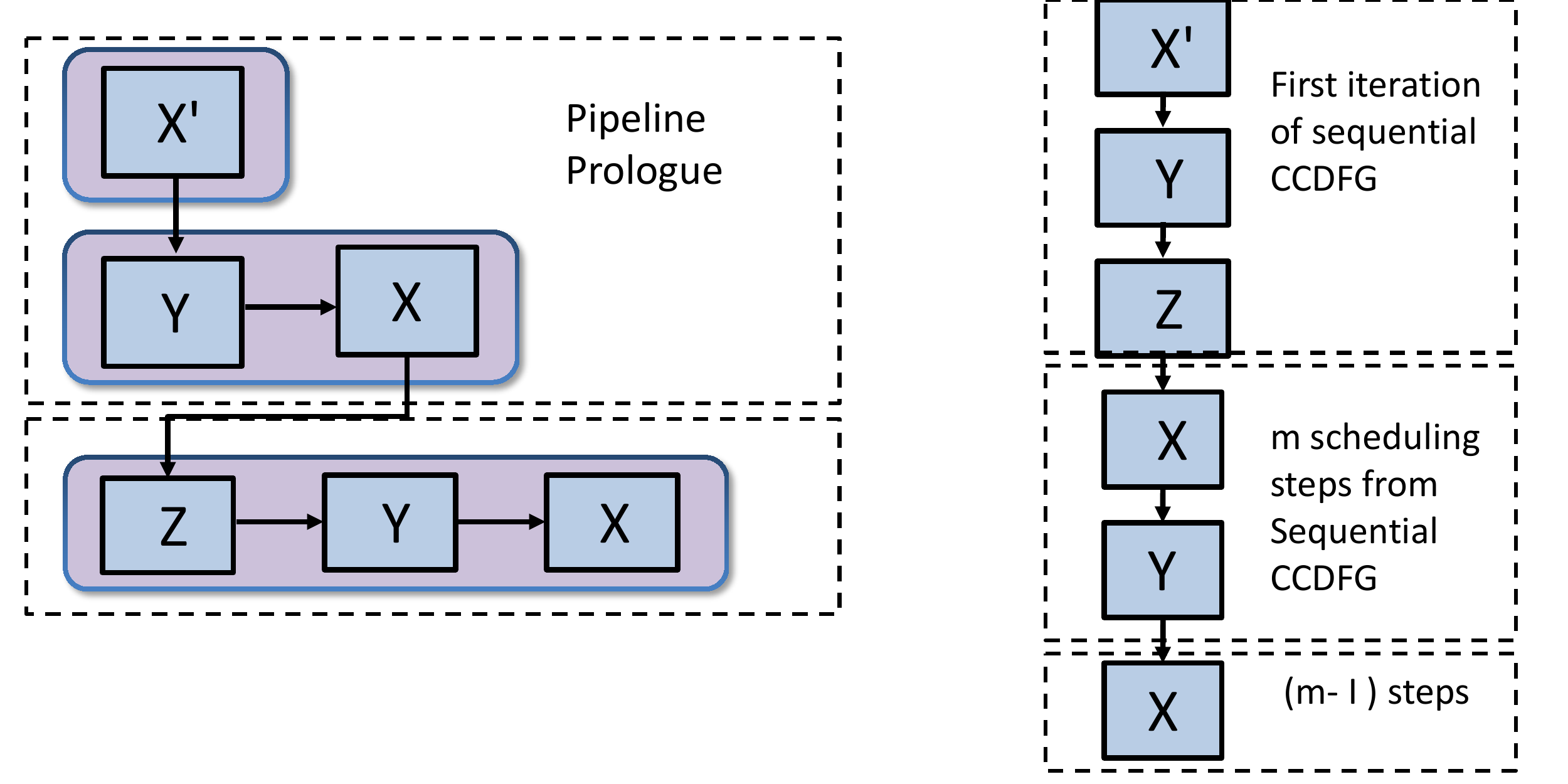}
&
\includegraphics[height=2.2in]{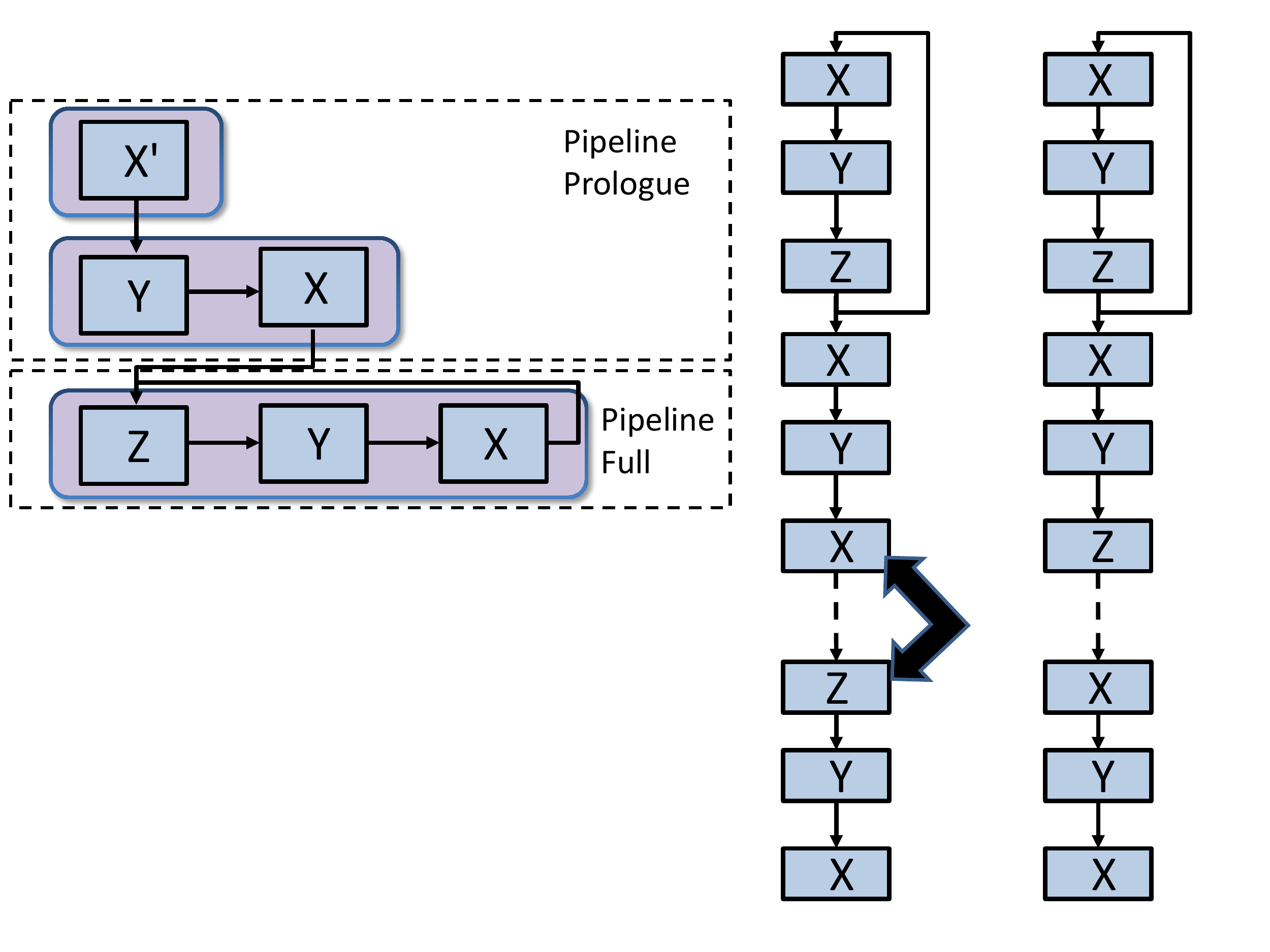}
\\
(a) & (b)
\end{tabular}
\end{center}
\caption{(a) Invariant base case where $k = 1$, executing
  pipeline prologue and one pipeline full stage is the same
  as executing seq-pre followed by sequence of partially
  completed sequential loop CCDFG.  (b) Assuming that
  invariant is true for $k$ steps, executing one pipeline full
  stage on both sides gives us $(k + 1)$ iterations of
  sequantial loop CCDFG followed by partially completed
  sequences as expected.}
\label{fig:invariant-proof}
\end{figure}

Although these requirements justify that our correspondence
relation is an invariant, they are used somewhat differently
in the base case (when the number of iterations $k$ of the
pipelined loop is $1$) and inductive step (assume the
invariant holds for $k$ iterations of the pipeline and prove
that it holds for $(k+ 1)$ iterations).  Their usage is
pictorially shown in Figure~\ref{fig:invariant-proof}.  For
the base case, we commute operations in the loop prologue of
the pipeline (which corresponds to the first iteration after
unrolling) with the loop body, while for the inductive step
we work with two consecutive iterations of the loop.

We have formalized the
correspondence-relation.  We have also proved the
implication chain from this relation to the correctness
statement and made progress in the proof of invariance (but
not completed it yet).  Our current ACL2 script has $125$
definitions and $150$ lemmas, including many lemmas about
structural properties of CCDFGs (but not counting the false
starts discussed in Section~\ref{sec:user-experience}).  We
are still working on proving some of the key components of
the proof obligations to show that the algorithm does not
introduce data hazards.

\section{Approaches, False Starts, and ACL2 Experience of a New User}
\label{sec:user-experience}

The invariant discussed in the preceding section, even if
complex, is somewhat natural in hindsight.  However, it is
actually the result of several iterations and false starts.
In this section we discuss some of these, and also touch
upon our experience with using ACL2 for a proof like this,
particularly from a new user's viewpoint (see below).  We
explain these experiences with the hope that they may help
other new users attempting something similar to avoid some
of the pitfalls.

Our invariant is very different from a typical invariant
used in the verification of pipelined machines (\eg, for
microprocessor pipelines).  We make explicit the
correspondence with the sequential execution.  The key
requirement from a pipeline invariant, \viz, hazard freedom,
is left implicit and arises indirectly as a proof obligation
for invariance of this predicate.  Most microprocessor
pipeline verification work went the other way.  For
instance, Sawada and Hunt's invariant~\cite{sh:pipeline},
expressed through an intermediate structure called MAETT,
``tracks'' the instructions as they pass through different
pipeline stages to ensure that hazards are not introduced.
One difference in our case is that we are not working with a
concrete pipeline with a fixed set of operations but an
algorithm that generates pipelines with an arbitrary
sequence of scheduling steps; a construction like MAETT is
thus not directly applicable.  However, there is a deeper
reason for defining our invariant the way we did.  Consider
the pipeline in Figure~\ref{fig:invariant-intro}.  Ignoring
the back edge (\eg, the edge from $X$ to $Z$ in the
scheduling step marked ``pipeline full'' in the pipelined
CCDFG), three columns in Figure~\ref{fig:invariant-intro}(b)
correspond roughly to three iterations of the sequential
loop.  Suppose we simply unroll the loop in the sequential
design three times, and then use a technique similar to
MAETT to track scheduling steps in this ``unrolled loop
body'' in the pipeline execution.  Unfortunately, this does
not work, because of the back edge.  There is no direct
correlation between this edge and any edge in the sequential
loop.  In fact, it is interesting to observe what its
introduction achieves: completion of one scheduling step in
each of the three partially executed, overlapping loop
iterations.  This suggests that the invariant must
explicitly capture the state of the executions that have
been partially completed during each iteration of the
pipeline (\ie, each traversal of the back edge).

The above is reflective of one of our major false starts: in
our initial approach we had decided to simplify the problem
by ignoring the back edge and proving the correspondence
between an unrolled loop and the pipeline as outlined above.
Only after substantially completing this proof and in
attempting to extend it to the pipeline with the back edge
did we realize that the extension does not work.

Another false start was in the way we initially formalized
hazard freedom.  Hazard freedom entails showing the
following. ``Suppose a variable $v$ is written by a
scheduling step $S$ and read subsequently by a scheduling
step $S'$ in the sequential CCDFG.  Then in the pipelined
CCDFG, there is no scheduling step $P$ that writes $v$ and
is executed between $S$ and $S'$.''  Originally, we defined
this notion directly for each variable, \viz, with a
function that statically analyzes the CCDFG to identify the
range of scheduling steps between a write and subsequent
read of each variable.  However, this does not directly
work.  For example, proving this property for variable $x$
may require a similar property to hold for another variable
$y$ (perhaps because $x$ is assigned an expression involving
$y$).  But the range of scheduling steps in which $x$ and
$y$ are read and written are different, and the extension of
the property to all the variables cannot be easily specified
by an invariant for any specific scheduling step.  Our
current approach alleviates this problem in that it
succinctly captures an ``on-track property'', \viz, that the
state after $k$ pipeline iterations is equivalent to partial
execution of a certain number of iterations in the
sequential CCDFG (in addition to completion of $k'$
iterations) which avoids this problem and can indeed be
specified as an invariant.

Perhaps the overall lesson is that it is foolhardy to start
mechanization until sufficient outline is completed as hand
proof.  However, it is impossible to fill in everything with
a hand proof.  Spotting whether an outline is sufficient
requires significant maturity, both with reasoning and with
the limits of what is doable automatically with ACL2.  Of
course, this will certainly always remain a creative
process.  However, published ACL2 proofs (in the workshop
and elsewhere) rarely touch upon false starts, focusing only
on the key steps of successful proofs.  It is difficult to
understand from that material at which level the authors
understood the proof before embarking on ACL2 and what paths
did or did not work.  Our experience directly supports
Moore's appeal~\cite{moore-fmcad-tutorial} to highlight
``the `minor' decisions that represent major breakthroughs
in the problem'' at least in the context of describing
successful (or failed) proofs.

The above pitfalls may be a by-product of the team involved
in this effort.  Of the four authors listed, three have had
no previous exposure to theorem proving; the remaining
author participated in the project in a consulting and
advisory role --- helping with questions at a high level and
with generic ACL2 proofs.  The bulk of the ACL2 work was
done by the first author, to whom this was the first project
beyond exercises in the ACL2 book.  This is clearly not a
typical ACL2 project team doing proofs at this scale.
Indeed, ACL2 is not designed for effective usage of teams
with this level of expertise.  For instance, the ACL2
textbook~\cite{car} states the following. 

\begin{quote} 
It takes weeks or months to learn to use the language and
the theorem prover, but months to years to become really
good at it.
\end{quote}

It also advises the user to start one's ACL2 project with
toy exercises.  While this is good advice, there are
situations where one cannot afford either a large team with
ACL2 expertise or the luxury to spend months to years to
become proficient ACL2 users.  Consequently, we believe it
is critical to find ways to address the question of learning
curve in theorem proving in order for ACL2 to make effective
forays in industrial contexts beyond organizations where it
has already been established.  We hope that our experience,
including the pitfalls mentioned above, will serve as a
basis for rethinking and conversation in the ACL2 community
on possible self-teaching materials to enable fast ramp-up
on ACL2 expertise for industrial and industrial-strength
projects.

We end this section with discussion with a few notes on
limitations in ACL2 documentation that, although simple once
resolved, stumped a new ACL2 user.  These are taken directly
from the records of the first author during the initial
phases of this work as a new user.  ACL2 is a complex
system.  The amount of documentation it provides is
impressive, and in our experience, far beyond that by most
other tools of similar complexity.  Nevertheless, as is
expected of a tool with more than two decades of
development, there are places where documentation is not
complete nor at the appropriate level for a new user.  It is
our endeavor to point out these corner cases in the hopes
that this will improve the documentation further, not to
complain about limitations.  We hope it will be taken in
that spirit.  Of course, the documentation has certainly
evolved and improved since these notes were taken.  We did
not check if the current documentation would have addressed
the confusion better, since what makes sense as a more
experienced user today might not have made sense to a new
user.

%% We spotted four key problems with the documentation.
%% \begin{enumerate}
%% \item Some Common Lisp primitives such as {\tt endp}
%% \item Discussion on termination
%% \item Formalization of induction hints
%% \item String manipulation functions
%% \end{enumerate}

\paragraph{Commonly used primitives link {\tt endp}:}  
The first author wanted to understand its usage after seeing
some definitions written by others, that used this
construct.  However, {\tt :DOC endp} merely suggests that it
is the same as {\tt atom} (a function unfamiliar to her),
and then discusses the difference in their guards; {\tt :DOC
  atom} suggests it is not a {\tt cons} (together with a
discussion of guards again).  In addition to not providing
any explanation of the usage of {\tt endp}, this discussion
inadvertently seemed to suggest that one must understand the
lengthy documentation on guard to make further progress.
Perhaps a small explanation that {\tt endp} is used as a
base case of functions that recur over a list using {\tt
  cdr} (\eg, when the list when a list is empty) would have
dispelled this confusion early.

\paragraph{Measure and Termination:}   
Most of our functions recur on complex graph structures and
one must supply a measure.  Unfortunately, documentation at
the new user level is lacking.  {\tt :DOC measure} merely
points to {\tt :DOC xargs}.  The latter provides an
intimidating example that involves all possible xargs (and
that the documentation admits is nonsensical).  Delving into
the {\tt :MEASURE} argument, one reads the following.

\begin{quote}
Value is a term involving only the formals of the function
being defined.  This term is indicates what is getting
smaller in the recursion. The well-founded relation with
which successive measures are compared is {\tt o<}. Also
allowed is a special case...
\end{quote}

\noindent
What it does not say is what we need to actually put in the
measure, and what we need to prove about it.  {\tt :DOC
  defun} provides a bit more explanation but also with no
examples, and the discussion here also seemed to indicate
(at least to the author) that one has to understand ordinals
at some depth before attempting to provide measure.

\paragraph{Induction Hint:}  
Documentation of induction hint suffers from similar
problem.  {\tt :DOC hint}, the topic where one is led from
an Internet search for induction hint in ACL2, again
provides an example of form of all possible hints, but with
little insight on what to write for induction hint and how
it will work.  The subtopic {\tt :induct} provides some clue
on what to do, \eg, it says the following:

\begin{quote}
...if value is a term other than t, then not only should the
system apply induction immediately, but it should analyze
value rather than the goal to generate its induction
scheme. Merging and the other induction heuristics are
applied....
\end{quote}

Admittedly, the first sentence above certainly provides a
complete description of what one needs to know to give an
induction hint, \viz, that the term provided will be
analyzed to generate the scheme.  However, this cryptic
explanation is difficult to translate to an actionable item
for actually generating induction.  An example where a
non-trivial induction hint is used would have helped the
user understand how to use it.

\paragraph{String Manipulation Functions:}  
Our project makes significant use of functions for string
manipulation.  However, finding them in the documentation is
a challenge.  For example, a simple Internet search on
``ACL2 functions for converting strings to symbols'' did not
turn up with {\tt packn}, resulting in a number of
unsuccessful attempts at defining something equivalent until
its existence was pointed to by a more experienced member.

\bigskip

The above is perhaps indicative of a more general problem.
ACL2 documentation is vast and complicated, and it is easy
to get lost.  We typically used Internet search to find a
topic of interest.  However, this requires a significant,
searchable (by Internet search) description of the different
available function symbols.  Many of these functions may
have been developed by the community and may not have
adequate descriptions provided, making this a challenge.
Perhaps the new initiative of ``Combined ACL2 + Community
Books Manual'' is a step in this direction, if the result
becomes searchable via standard Web search.

\bigskip

Of course, documentation suggestions necessarily are narrow
and low-level, and one might ask why we bother to point them
out in a paper rather than through personal communication
with the ACL2 authors, or even try to fix them ourselves in
the ACL2+books combined manual which is maintained and
developed by the community rather than the authors of
ACL2.\footnote{After the acceptance of this paper, Kaufmann
  requested us to make some of the changes to the
  documentation to address the comments below, and the first
  author has volunteered to do so.}  The reason we
nevertheless point them out in the paper is that they are
symptomatic of a larger issue which we believe merits deeper
discussion, deliberation, and collaboration of the entire
community.  As pointed out above, the standard model of
successful application of ACL2 in the industrial projects
has been through the efforts of one or more experts with
years of experience in the theorem prover working full-time
with ACL2 on an industrial application.  However, our
experience suggests that this large barrier to entry often
dooms the possibility of a new ``customer'' looking into it
as a solution, long before it has had a chance to ``prove
its mettle'' on the customer's problem.  Addressing this
requires us to rethink how we can make ACL2 useful to
promising but new users taking on an industrial project, at
least at the level in which a new user can take on other
formal verification and validation tools.  Our experience
suggests that if that were possible, there would be several
new industrial partners willing to try it out.  However,
achieving that does require the community collaboratively
rethinking how the theorem prover ought to be presented to
such a new user.  We do not have a solution to offer, but we
hope that our remarks would cause some deliberation and
discussion within the community.  Identifying documentation
pitfalls is a tiny contribution to this effort from our
part.

\section{Conclusion}
\label{sec:concl}

There has been a significant amount of work on pipeline
verification, both within and outside the ACL2
community~\cite{bd:pipeline,sh:pipeline,pm:pipelines,velev05}.
However, most of pipeline verification research has focused
on architectural pipelines, in particular pipelined
microprocessors.  There are significant differences in goals
and techniques between these efforts and ours.
Microprocessor pipelines include optimized (hand-crafted)
control and forwarding logics, but a static set of
operations based on the instruction set.  Loop pipelines
tend to be deep with a high complexity at each stage, but
control and forwarding logics are more standardized since
they are automatically synthesized.  
%It is hard to adopt SEC
%techniques for microprocessor pipelines directly for
%function pipelining, \eg, lack of a standardized instruction
%set makes it difficult to identify targets for uninterpreted
%functions. 
Furthermore, microprocessor pipeline verification
is focused on one (hand-crafted) pipeline implementation,
while our work focuses on verifying an {\em algorithm
  producing pipelines}.  As we discussed in
Section~\ref{sec:user-experience}, abstraction techniques
such as MAETT~\cite{sh:pipeline} does not apply to our case
and we had to come up with a very different invariant.

Our work is very closely related to recent work on
verification of software pipelines.  In particular, Tristan
and Leroy~\cite{tl:software-popl10} present a verified
translation validator for software loop pipelines.  The loop
pipelines in behavioral synthesis considered in this paper
are close in structure to software loop pipelines, although
our formalization (\eg, CCDFG) has different semantics from
the Control Flow Graphs they use, reflecting the difference
between eventual targets of compilation (\viz, hardware
vs. software).  However, the fundamental difference is in
the approach taken to actually certify the pipelines.
Tristan and Leroy's approach decomposes the certification
problem into two parts, a ``dynamic'' part that is certified
on a case-by-case basis  and a ``static'' part that is
certified in the Coq theorem prover once and for all.  The
theorem proven by Coq is informally paraphrased as follows:

\begin{quote}
Suppose the pipelining algorithm generates a pipeline
${\cal{P}}$ from a sequential design ${\cal{S}}$.  Suppose
symbolic simulation of ${\cal{S}}$ and ${\cal{P}}$ verifies
certain ``dynamic'' verification conditions (VCs).  Then
${\cal{S}}$ and ${\cal{P}}$ are indeed semantically
equivalent.
\end{quote}

\noindent
Thus for any pipeline instance ${\cal{P}}$ generated by
their algorithm, symbolic simulation is executed between
${\cal{P}}$ and ${\cal{S}}$ to certify that ${\cal{P}}$ is
indeed a correct pipelined implementation of ${\cal{S}}$.
The dynamic VCs checked by symbolic simulation essentially
certify that the pipeline generation did not overlook any
hazards.

This is where our work differs from theirs.  Our work is
expected to provide a single theorem certifying the
correctness of the reference pipelined implementation,
without requiring further runtime hazard check.
Furthermore, their correspondence theorem relates the
pipelined implementation with a sequential design with a
(bounded) unrolled loop, while our approach certifies the
correspondence between the actual Control Flow Graph (CFG)
and the pipelined implementation.  Indeed, Tristan and Leroy
remark that the mechanization of the correspondence between
the CFG and unrolled loop is ``infuriatingly difficult''.
We speculate this is so because they focus on verifying the
correspondence between the unrolled loop and the pipeline.
As we discuss in Section~\ref{sec:user-experience},
attempting the formal correspondence between the unrolled
sequential loop and pipelined design was not possible since
there is no formal way to connect to back edge of the loop
with any of the edges in the pipeline.  We believe that
reconciling this problem and developing a fully certified
pipeline generation algorithm would require backtracking
from the correspondence with an unrolled loop (and hence
translation validation) to a more complex invariant like
ours.  Of course we must note that we can ``afford'' to
develop a fully certified algorithm in our approach since
the pipelines are simpler
(cf. Section~\ref{sec:formalization}); achieving this for
arbitrary software pipeline may require further more subtle
invariants.

In addition to technical contributions, we see our work as
providing an important methodological contribution enabling
use of theorem proving in situations where one needs to
certify the result of an implementation on which theorem
proving cannot be directly applied either because it is
closed-source or because it is highly complex: (1)~create a
reference implementation, perhaps using as much information
as available from the actual implementation, in our case
information about pipeline intervals, (2)~certify this
simpler reference implementation with theorem proving, and
(3)~develop an SEC framework to compare the result of the
reference implementation with that of the actual
implementation.  In addition to making theorem proving
applicable on industrial flows without requiring us to
certify industial implementations with their full
complexity, this approach permits adjusting the algorithm
(within limits) to suit mechanical reasoning while still
affording comparison with actual synthesized artifacts.  We
have made liberal use of this ``luxury'', \eg, we are
continually redefining our superstep construction function
to facilitate proof of key structural lemmas of the
invariant.  We believe similar approach is applicable in
other contexts and may provide effective use of theorem
proving within industrial verification flows.

\bibliographystyle{eptcs}
\bibliography{bib}

\end{document}